# Observation of Spin-Orbit Effects with Spin Rotation Symmetry

Alisha M. Humphries[1,♣], Tao Wang[2,♣], Eric R. J. Edwards[3], Shane R. Allen[1], Justin M. Shaw[3], Hans T. Nembach[3,4], John Q. Xiao[2], T. J. Silva[3], Xin Fan[1,*]

1. Department of Physics and Astronomy, University of Denver, Denver, CO, 80210, USA
2. Department of Physics and Astronomy, University of Delaware, Newark, DE, 19716, USA
3. Quantum Electromagnetics Division, National Institute of Standards and Technology, Boulder, CO, 80305, USA
4. JILA, University of Colorado, Boulder, Colorado 80309, USA

[♣] These authors contribute equally.

* xin.fan@du.edu

The spin-orbit interaction enables interconversion between a charge current and a spin current. It is usually believed that in a nonmagnetic metal (NM) or at a NM/ferromagnetic metal (FM) bilayer interface, the symmetry of spin-orbit effects (SOE) requires that the spin current, charge current and spin orientation are all orthogonal to each other. Here we show the observation of a SOE near the NM/FM interface that exhibits a very different symmetry from the conventional spin Hall effect, insofar as the spin polarization is further rotated about the magnetization. These results imply that a perpendicularly polarized spin current can be generated with an in-plane charge current simply by use of a FM/NM bilayer with magnetization collinear to the charge current. The ability to generate a spin current with arbitrary polarization using typical magnetic materials will greatly benefit the development of magnetic memories.

The spin-orbit interaction enables interconversion between a charge current and a spin current[1-10]. It has been shown that an in-plane charge current in a FM/NM bilayer can generate spin-orbit torques (SOT) via the bulk spin Hall effect in the NM[7] and/or from the interfacial SOEs at the FM/NM interface[11-13]. These effects can be used for magnetization switching with an in-plane charge current, with potential benefits for the development in magnetic random access memories (MRAM)[14]. The symmetry of the spin current generation for the spin Hall effect is captured by the essential phenomenology,

$$\vec{Q}_{\hat{\sigma}} = \frac{\hbar}{2e}\theta \vec{j}_e \times \hat{\sigma}, \qquad (1)$$

where $\vec{j}_e$ is the in-plane charge current density, $\vec{Q}_{\hat{\sigma}}$ is the out-of-plane flowing spin current density, where the subscript denotes its spin polarization $\hat{\sigma}$, $\theta$ is the spin/charge conversion efficiency, $\hbar$ is the reduced Planck's constant, and $e$ is the electron charge. According to Eq. (1), an in-plane charge current can generate an out-of-plane flowing spin current, but only with spins polarized in-plane and perpendicular to the charge current. As such, the direct switching of a perpendicular magnetized film via the combination of the spin Hall effect and spin torque transfer (i.e. anti-damping) is not possible. To cause such switching, additional sources of broken symmetry are required, such as an intrinsic gradient of the magnetic anisotropy[15], tilting of the magnetization by an external magnetic field relative to the interface normal[16,17], or an effective exchange field[18,19]. Even then, if tilting of the magnetization facilitates the switching process, the SOT must necessarily overcome both the torque due to anisotropy as well as that of the damping. As such, the efficiency of such a switching process is necessarily compromised.

A spin current with an *out-of-plane polarization* can switch perpendicular magnetization via the anti-damping process without the need to tilt the magnetization. Presumably additional symmetry breaking is required for this to happen. For example, MacNeill *et al.*[20] recently showed that a spin current with unconventional symmetry can be generated in a WTe$_2$/Permalloy bilayer due to the unique crystal symmetry of the transition-metal dichalcogenide. Alternatively, Taniguchi *et al.*[21] have proposed that an out-of-plane polarized spin current can be generated via the combination of the anomalous Hall effect in a FM with tilted magnetization and the spin filtering effect. More generally, Amin and Stiles have predicted that spin-orbit scattering of an in-plane charge current at a FM/NM interface can give rise to a spin current with an arbitrary spin polarization, because of the interaction between spins and the magnetic order at the interface[22]. One possible microscopic mechanism consistent with such a prediction is the case where spin polarization of a spin current generated near the FM/NM interface precesses about the magnetization. Although transverse spins rapidly dephase in a FM[23,24], this is not necessarily the case at the FM/NM interface or when FM is very thin. Therefore, from a purely phenomenological point of view, we might expect a source of spin current described by

$$\vec{Q}_{\hat{\sigma}}^{R} = \frac{\hbar}{2e}\theta^{R}\vec{j}_{e}\times(\hat{m}\times\hat{\sigma}) \qquad (2)$$

where $\theta^R$ is the spin/charge conversion efficiency for the SOE with the rotated spin symmetry. In this sense, the generation of a spin current described by Eq. (2) is loosely analogous to the rotation of the polarization of light by the Faraday effect.

As shown in Fig. 1(a), when an in-plane charge current passes through a FM/NM interface, out-of-plane spin currents can be generated in accordance with both Eqs. (1) and (2). It should be emphasized that Eq. (2) describes an effect that is inherently different from the spin filtering proposed by Taniguchi et al.[21]. The polarization of the spin current generated via spin filtering is always polarized collinear with the magnetization, whereas the spin current due to spin rotation is always polarized orthogonal to the magnetization.

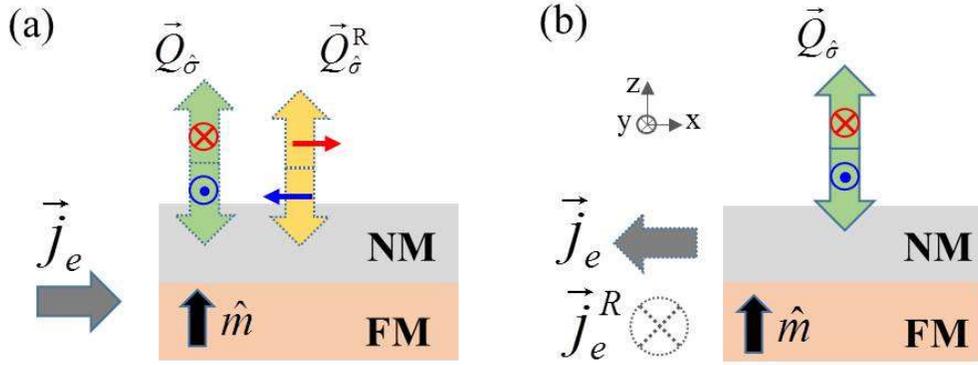

Figure 1 (a) A sketch to illustrate the spin currents generated from a charge current near the FM/NM interface. The red and blue arrows represent spins. The green and yellow arrows represent the spin current with conventional symmetry, $\vec{Q}_{\hat{\sigma}}$, and the spin current with spin rotation symmetry, $\vec{Q}_{\hat{\sigma}}^{R}$, respectively. The grey arrow represents the charge current $\vec{j}_e$. (b) A sketch of the reciprocal process to illustrate how the conventional and rotated charge currents are generated by a pure spin current near the FM/NM interface.

Similarly, as illustrated in Fig. 1(b), a spin current $\vec{Q}_{\hat{\sigma}}$ that flows out-of-plane in a FM/NM bilayer can generate two in-plane charge currents via the spin galvanic effect (SGE), one with $\vec{j}_e$ in the direction $\vec{Q}_{\hat{\sigma}}\times\hat{\sigma}$, and one with $\vec{j}_e^R$ in the rotated direction $-\vec{Q}_{\hat{\sigma}}\times(\hat{m}\times\hat{\sigma})$. This process can be mathematically described as

$$\begin{cases} \vec{j}_e = \frac{2e}{\hbar}\theta\vec{Q}_{\hat{\sigma}}\times\hat{\sigma} \\ \vec{j}_e^R = -\frac{2e}{\hbar}\theta^R\vec{Q}_{\hat{\sigma}}\times(\hat{m}\times\hat{\sigma}) \end{cases}, \qquad (3)$$

where the negative sign in the second equation is necessary to satisfy the Onsager relation as discussed in the Supplementary Information 1.

**Results**

Here we show experimental observations of a SOE with spin rotation symmetry by use of current-induced SOT and spin Seebeck effect (SSE)-driven SGE measurements.

First, we present the detection of the spin current with rotated spins generated near an interface between Cu and a perpendicular magnetized layer (PML), as described by Eq. (2). The test sample is a multilayer with the structure seed/PML/Cu(3)/Py(2)/Pt(3), and the control sample has the structure seed/PML/Cu(3)/TaO$_x$(3)/Py(2)/Pt(3), where seed = Ta(2)/Cu(3), PML = [Co$_{90}$Fe$_{10}$(0.16)/Ni(0.6)]$_8$/Co$_{90}$Fe$_{10}$(0.16), Py = Ni$_{80}$Fe$_{20}$, and the numbers in parentheses are nominal thicknesses in nanometers. The Py layer is the spin current detector. The TaO$_x$ insulating layer in the control sample blocks the flow of spin current between the PML and Py layers. The electrical and magnetic properties of the test sample are shown in the Supplementary Information 2.

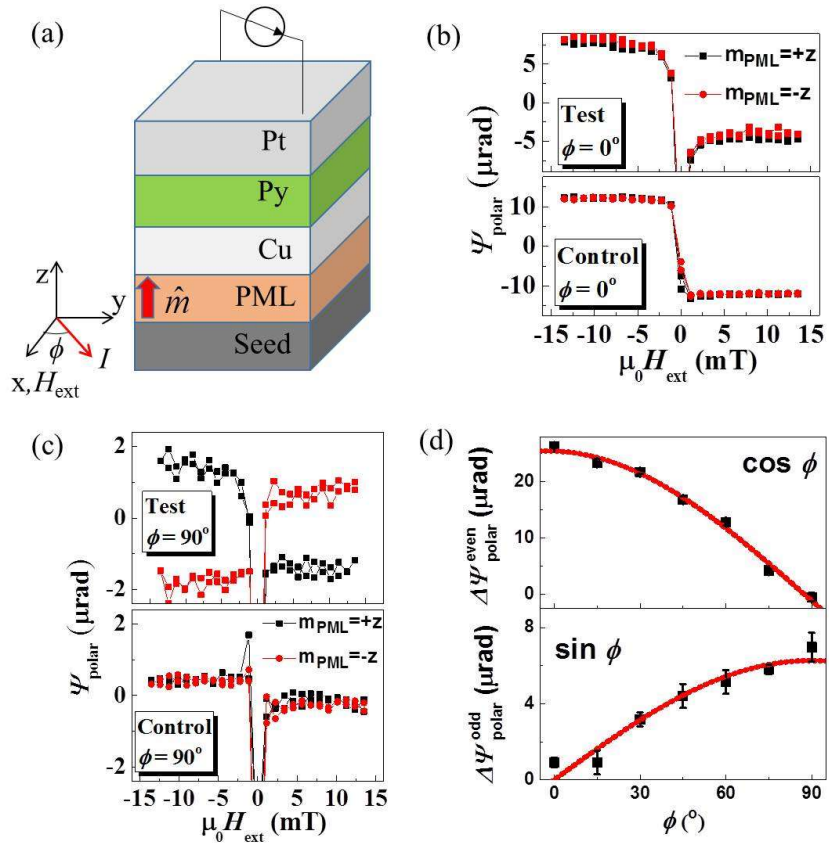

Figure 2 (a) Experimental configurations for the SOTs measured by MOKE. In this measurement, an external field is swept along the x-direction. The sample together with the applied current can be rotated in the film plane. The out-of-plane magnetization tilting in Py due to damping-like torques is measured by polar MOKE, in which the light is incident normally with linear polarization 45° from the x-direction. (b) The polar MOKE response measured in the test (top) and control samples (bottom) when current is applied parallel with $H_{\text{ext}}$. No dependence on $\hat{m}$ is observed. (c) The polar MOKE response measured in the test (top) and control samples (bottom) when current is applied perpendicular to $H_{\text{ext}}$. In the test sample, the polar MOKE response is reversed when $\hat{m}$ is reversed. In contrast, the polar MOKE response in the control sample has little dependence on $\hat{m}$. The weak hysteresis-like signal in the control sample is likely due to small misalignment of $H_{\text{ext}}$. (d) The angle dependence of $\Delta\psi_{\text{polar}}^{\text{even}}$ and $\Delta\psi_{\text{polar}}^{\text{odd}}$ of the test sample. Here we define $\Delta\psi_{\text{polar}}^{\text{even}} = \left(\psi_{\text{polar}}^{++} - \psi_{\text{polar}}^{+-}\right) + \left(\psi_{\text{polar}}^{-+} - \psi_{\text{polar}}^{--}\right)$ and $\Delta\psi_{\text{polar}}^{\text{odd}} = \left(\psi_{\text{polar}}^{++} - \psi_{\text{polar}}^{+-}\right) - \left(\psi_{\text{polar}}^{-+} - \psi_{\text{polar}}^{--}\right)$, where the first superscript in $\psi_{\text{Polar}}^{++}$ denotes the sign of $\hat{m}$ and the second superscript denotes the sign of $\hat{m}_{\text{Py}}$ during the measurement. Values when $\hat{m}_{\text{Py}}$ is saturated are used in the extrapolation. The red lines are fittings using $\cos\phi$ and $\sin\phi$ functions.

The measurement geometry is shown in Fig. 2(a). According to Eq. (2), an in-plane charge current $\vec{j}_e$ generates two spin currents that exert torques on the Py magnetization $\hat{m}_{\text{Py}}$: $\vec{Q}_{\hat{\sigma}}$ with $\hat{\sigma} \parallel (\vec{j}_e \times \hat{z})$ due to the SOE with conventional symmetry near the Pt/Py and PML/Cu interfaces, and $\vec{Q}_{\hat{\sigma}}^{\text{R}}$ with $\hat{\sigma} \parallel \hat{m} \times (\vec{j}_e \times \hat{z})$ due to the SOE with spin rotation symmetry. Here, $\hat{m}$ is the unit magnetization vector of the PML. Since the spin polarizations of $\vec{Q}_{\hat{\sigma}}$ and $\vec{Q}_{\hat{\sigma}}^{\text{R}}$ are both in-plane, the effect of the damping-like SOT from both spin currents is to tilt $\hat{m}_{\text{Py}}$ out-of-plane[25]. In the limit where the current-induced SOT is small relative to the torque due to the applied field and the demagnetizing field, the out-of-plane component of the Py magnetization is given by

$$m_{\text{Py}}^{\perp} = \frac{h_{\text{DL}}\left[\hat{m}_{\text{Py}} \times \left(\hat{j}_e \times \hat{z}\right)\right] \cdot \hat{z} + h_{\text{DL}}^{\text{R}}\left[\hat{m}_{\text{Py}} \times \left(\hat{m} \times \left(\hat{j}_e \times \hat{z}\right)\right)\right] \cdot \hat{z} + h_{\text{Oe}}^{\perp}}{|H_{\text{ext}}| + M_{\text{eff}}}, \qquad (4)$$

where $\hat{j}_e$ is the unit vector along $\vec{j}_e$, $M_{\text{eff}}$ is the effective Py demagnetizing field along the z-direction, $h_{\text{Oe}}^{\perp}$ is the current-induced out-of-plane magnetic field due to the Oersted field, $h_{\text{DL}}$ and $h_{\text{DL}}^{\text{R}}$ are the equivalent fields due to the damping-like (DL) SOTs generated by $\vec{Q}_{\hat{\sigma}}$ and $\vec{Q}_{\hat{\sigma}}^{\text{R}}$, respectively. A more thorough analysis of spin-orbit torques is shown in the Supplementary Information 3.

We detect $m_{\text{Py}}^{\perp}$ by use of the polar magneto-optic-Kerr-effect (MOKE)[25], which results in the polarization rotation $\Psi_{\text{polar}}$ of linearly polarized incident light. The three terms in Eq. (4) can be distinguished by their dependence on $\hat{m}_{\text{Py}}$ and $\hat{m}$. In the measurement geometry with $\phi = 0°$, the second term in Eq. (4) is zero. As shown in Fig. 2(b), signals of $\Psi_{\text{polar}}$ in both the test and control samples resemble the Py magnetization hysteresis, which can be understood from the first term in Eq. (4). The signal is independent of $\hat{m}$. By performing a linescan measurement[26], we estimate that $h_{DL}$ is about $120 \pm 12$ A/m in the test sample, when the total integrated current density across the entire film is 600 A/m.

In the measurement geometry with $\phi = 90°$, the first term in Eq. (4) is zero. As shown in Fig. 2 (c), $\Psi_{polar}$ for the test sample switches with the applied field direction, and also reverses polarity when $\hat{m}$ is switched, which is consistent with the behavior expected from the second term of Eq. (4). The magnitude of $h_{DL}^{R}$ is estimated to be 25% of the magnitude of $h_{DL}$, or $30 \pm 4$ A/m for a current density of $3.8 \times 10^{10}$ A/m² in the PML. This result confirms the generation of a spin current with rotated spin polarization by the PML. By use of Eq. (2), we estimate

$$\theta^R = \frac{2e}{\hbar} \frac{|\vec{Q}_{\hat{\sigma}}^R|}{|\vec{j}_e|} = \frac{2e}{\hbar} \frac{\mu_0 M_{Py} d_{Py} h_{DL}^R}{|\vec{j}_e|} \approx (4.8 \pm 0.6) \times 10^{-3},$$

under the assumption of perfect spin absorption at the Py/Cu interface, where $\mu_0 M_{Py} = 1$ T and $d_{Py} = 2$ nm are the saturation magnetization and thickness of Py, respectively. For the control sample, where $\vec{Q}_{\hat{\sigma}}^R$ is presumably suppressed by the TaO$_x$ layer, $\Psi_{polar}$ is independent of $\hat{m}$. The slight dependence of $\Psi_{polar}$ on the applied field is possibly due to misalignment of the applied field and the current flow direction.

We decomposed $\Psi_{polar}$ into the component that is even in $\hat{m}$ ($\Delta\psi_{polar}^{even}$) and odd in $\hat{m}$ ($\Delta\psi_{polar}^{odd}$), then measured the dependence of the two components on the applied field angle in the sample plane, where $\phi$ is the angle between the applied field and the charge current direction. As shown in Fig. 2 (d), $\Delta\psi_{polar}^{even}$ is proportional to $\cos(\phi)$, whereas $\Delta\psi_{polar}^{odd}$ is proportional to $\sin(\phi)$, consistent with the phenomenology expressed in Eq. (4).

We also measured the current-induced field-like torques for the same sample. As shown in the Supplementary Information 4, we observed the same dependencies on $\hat{m}$ as for the damping-like torque. Other possible mechanisms leading to the observed signals are discussed and ruled out in the Supplementary Information 5-7.

To further validate our findings, we also measured the spin rotation symmetry of the SGE, described in Fig. 1 (b), with a SSE-driven SGE measurement of the same samples. As shown in Fig. 3 (a), when the samples are subject to an out-of-plane temperature gradient, a spin current is generated due to the SSE[27,28], which then generates an in-plane voltage. The voltage may arise from the anomalous Nernst effect in the magnetic layers, the SGE due to the spin currents injected into the adjacent layers, as well as the planar Nernst effect[29] in the PML. Depending on whether it has an even or odd symmetry with $\hat{m}$, the voltage can be described as

$$\begin{cases} V^{even} = \eta_{Py}\left(\vec{\nabla}T \times \hat{m}_{Py}\right)\cdot \hat{y} + \eta_{PML}\left(\vec{\nabla}T \times \hat{m}\right)\cdot \hat{y} \\ V^{odd} = \eta_R \vec{\nabla}T \times \left(\hat{m}_{Py} \times \hat{m}\right) + \eta_{PML}^{PNE}\left(\hat{m}\cdot \hat{y}\right)\left(\hat{m}\cdot \vec{\nabla}T\right) \end{cases}, \quad (5)$$

where $\vec{\nabla}T$ is the temperature gradient in the z-direction, $\eta_{Py}$, and $\eta_{PML}$, with units of V·m·K$^{-1}$, are the additive anomalous Nernst and SGE coefficients associated with the Py and PML layers, respectively, $\eta_{PML}^{PNE}$ is the coefficient associated with the planar Nernst effect of the PML, and $\eta_R$ is the coefficient associated to the SGE voltage with spin rotation symmetry described by the second equation of Eq. (3). Note that $\eta_R$ potentially has two competing sources: the spin current generated in Py that diffuses towards the PML, and the spin current generated in PML that diffuses towards the Py.

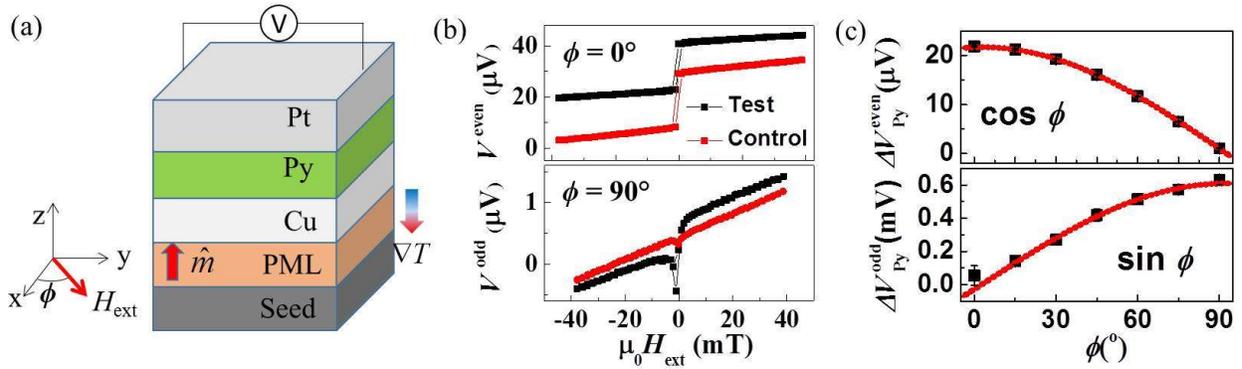

Figure 3 (a) Experimental configuration for the SSE-driven SGE measurements. (b) Voltages measured in the two different configurations for the test sample (seed/PML/Cu/Py/Pt) and control sample (seed/PML/Cu/TaO$_x$/Py/Pt). $V^{even}$ and $V^{odd}$ are the sum and difference of the two voltage curves when $\hat{m}$ is polarized up and down, respectively. (c) Angle dependence of the voltage signals associated with Py switching. Here $\Delta V_{Py}^{even}$ is the first term of $V^{even}$ in Eq. (5) and $\Delta V_{Py}^{odd}$ is the first term of $V^{odd}$ in Eq. (5). The red curves are fits to cos($\phi$) and sin($\phi$).

As shown in Fig. 3 (b), $V^{even}$ measured when $H_{ext}$ is along the x-direction consists of two components: one resembles the hysteretic switching of Py, and a linear slope related to the magnetization tilting of the PML under the influence of external field, as understood from Eq. (5). When $H_{ext}$ is applied along the y-direction, $V^{even}$ vanishes. $V^{odd}$ measured for the control sample yields a straight line, which is consistent with the planar Nernst effect described in Eq. (5). However, $V^{odd}$ measured for the test sample has an additional component related to the Py magnetization switching, which is consistent with the third term in Eq. (5) due to the SGE with spin rotation symmetry. Shown in Fig. 3(c), the angle dependences of the voltage signal further confirm the symmetry described by Eq. (5).

**Discussions**

It should be pointed out that the SOE with spin rotation symmetry may not only arise from the interface between the very top layer of PML and Cu. The PML consists of many interfaces of Ni/Co$_{90}$Fe$_{10}$, which are known to have a strong spin-orbit interaction that gives rise to the perpendicular anisotropy. Since each layer in the PML is very thin, the observed SOE may partially arise from the Ni/Co$_{90}$Fe$_{10}$ interfaces within the PML.

The SOE with spin rotation symmetry in combination with the SOE with conventional symmetry can generate spin current with arbitrary polarization simply by adjusting the magnetization direction. An important implication of these findings is the ability to generate a perpendicular polarized spin current by use of a FM/NM interface, where the FM is magnetized collinear to the current flow direction. Such a spin current polarization is required to switch a perpendicular magnetized layer by use of anti-damping spin transfer torque alone. These findings can significantly benefit the development of MRAM technology, where perpendicular magnetized memory layer is more favorable[30]. Although the verification of the spin rotation phenomenology presented here does not permit us to predict the efficiency of the perpendicularly polarized spin current generation for the case of collinear current and FM magnetization, we think the key to have high efficiency is through interface optimization, where spin-orbit interaction, spin precession and dephasing should all be taken into account.

## Methods

Sample Fabrication

The samples used in this study are fabricated by magnetron sputtering. The TaO$_x$ layers in the control samples are made by depositing 1.5 nm Ta film and subsequently exposing to the air. This process is repeated in order to fabricate a total of 3 nm TaO$_x$ layer.

MOKE Measurement

In the MOKE measurement, the sample is patterned into a 50 μm x 50 μm square. The total electric current applied is 30 mA, from which we estimate the current density through PML to be about 3.8 x 10$^{10}$ A/m$^2$. The principle of the SOT detection with MOKE and detailed protocols can be found in reference [25]. In the measurement, we apply a small in-plane sweeping external magnetic field $H_{ext}$ (<15mT) that aligns the magnetization of Py. Due to the large anisotropy (~390mT), the magnetization of the PML remains mostly perpendicular when $H_{ext}$ sweeps in the film plane. We set the initial magnetization direction of the PML by placing a permanent magnet close to the sample and then remove it. The permanent magnet generates about 50 mT field perpendicular to the film plane while the coercivity of the PML is about 30 mT. We typically measure the hysteretic loops for 15 times and take the average.

Thermal Measurement

In the thermal measurement, the samples are typically cut into 2 mm x 25 mm strips. The voltages across the samples are measured by a Keithley nano voltmeter 2182. The samples are sandwiched between two aluminum plates. The aluminum plates are attached to Peltier elements to create a temperature difference across the sample. The typical temperature difference, ΔT, measured on the two aluminum plates is about 50 K. All voltages are scaled to a 50 K temperature difference by taking $V/\Delta T \times 50$. Similar to the MOKE measurement, we switch the magnetization of PML by placing a permanent magnet close to the sample and then remove it. We measure the hysteretic loops 10 – 20 times and take the average. Possible drifts in the measurement are removed mathematically by assuming the drift is linear with measurement time.

## Acknowledgement

The work done at University of Denver is supported by PROF and Partners in Scholarship grant. The work done at University of Delaware is supported by NSF DMR1505192. The authors would like to thank Vivek Amin, Mark Stiles, Satoru Emori and Barry Zink for critical reading of the manuscript and illuminating discussions.

## Contributions

X.F. conceived and designed the experiments; E.E. and J.S. fabricated the samples; E.E. and S.A. characterized the samples; A.H., S.A. and X.F. performed the thermal measurement; T.W.

patterned the sample and performed the MOKE measurements; All authors contributed to analysis and interpretation of the data.

# Observation of Spin-Orbit Effects with Spin Rotation Symmetry

Alisha M. Humphries[1], Tao Wang[2], Eric R. J. Edwards[3], Shane R. Allen[1], Justin M. Shaw[3], Hans T. Nembach[3,4], John Q. Xiao[2], T. J. Silva[3], Xin Fan[1,*]

1. Department of Physics and Astronomy, University of Denver, Denver, CO, 80210, USA
2. Department of Physics and Astronomy, University of Delaware, Newark, DE, 19716, USA
3. Quantum Electromagnetics Division, National Institute of Standards and Technology, Boulder, CO, 80305, USA
4. JILA, University of Colorado, Boulder, Colorado 80309, USA

* xin.fan@du.edu

**Supplementary Information**

**Content**

1. **Time reversal symmetry of the spin/charge current conversion**

2. **Electrical and Magnetic properties of the sample**

3. **Analysis of current-induced magnetization reorientation**

4. **Field-like SOT with spin rotation symmetry**

5. **Discussion of an alternative mechanism for spin rotation**

6. **Artifacts due to the anomalous Hall effect**

7. **Artifacts due to the interlayer magneto static coupling**

1. **Time reversal symmetry of the spin/charge current interconversion**

   In this section we discuss why the negative sign is necessary in the second equation of Eq. (3) in the main text.

   Equations 1 and 2 in the main text describe that an in-plane charge current can generate two orthogonally polarized spin currents: $\vec{Q}_{\hat{\sigma}} = \frac{\hbar}{2e}\theta \vec{j}_e \times \hat{\sigma}$ with conventional symmetry and $\vec{Q}_{\hat{\sigma}}^R = \frac{\hbar}{2e}\theta^R \vec{j}_e \times (\hat{m} \times \hat{\sigma})$ with spin rotation symmetry. According to the Onsager relations, the time reversal process should also be valid. Under time reversal, both the spin current direction and its spin polarization reverses, that is $\vec{Q}_{\hat{\sigma}} \to -\vec{Q}_{\hat{\sigma}}$ and $\hat{\sigma} \to -\hat{\sigma}$. The charge current and magnetization are also odd under time reversal, $\vec{j}_e \to -\vec{j}_e$ and $\hat{m} \to -\hat{m}$. Therefore, in the reverse process with conventional symmetry, a spin current moving along $-\vec{Q}_{\hat{\sigma}}$ with spin polarization $-\hat{\sigma}$ can generate a charge current $-\vec{j}_e$. Such a process can therefore be written as $\vec{j}_e = \frac{2e}{\hbar}\vec{Q}_{\hat{\sigma}} \times \hat{\sigma}$. On the other hand, in the reverse process with spin rotation symmetry, a spin current moving along $-\vec{Q}_{\hat{\sigma}}$ with spin polarization $-\hat{\sigma}$ can generate a charge current $-\vec{j}_e^R$ in the presence of a magnetization along $-\hat{m}$. Therefore, such a reverse process can be written as $\vec{j}_e^R = -\frac{2e}{\hbar}\theta^R \vec{Q}_{\hat{\sigma}} \times (\hat{m} \times \hat{\sigma})$.

2. **Electrical and Magnetic properties of the sample**

   The sheet resistance of the test sample is measured to be about 8.8 Ω and the conductivity of the PML is about 7.1 x 10$^6$ Ω$^{-1}$m$^{-1}$.

   The magnetic hysteresis measurement of the test sample seed/PML/Cu(3)/Py(2)/Pt(3) is shown in Fig. S1. The remanence of the PML is almost 100%, and the magnetization tilting of the PML under a small in-plane external magnetic field is very small. From ferromagnetic resonance measurements, we determined the effective out-of-plane demagnetizing field of Py to be $\mu_0 M_{eff}$ = 0.74 T and the effective out-of-plane anisotropy field of the PML to be $\mu_0 H_{an\perp}$ = 0.39 T. In our measurements, the largest in-plane field applied is 40 mT, which tilts the PML magnetization by approximately 6°.

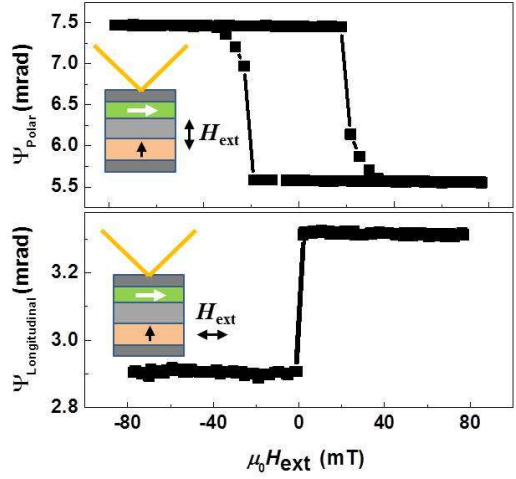

Figure S1. Magnetic hysteresis measured by MOKE with the external magnetic field swept out-of-plane and in-plane.

## 3. Analysis of current-induced magnetization reorientation

In the current-induced SOT measurement, besides the effective fields described in Eq. (4), an additional damping-like field may be generated by a component of the spin current emitted from the PML as a result of the spin filtering effect[1]. Such spin currents are necessarily polarized parallel to $\hat{m}$. In addition, there may be field-like torques generated by the spin currents. Here, we attempt to account for all possible fields as follows:

$$\begin{aligned}
\vec{H}_{tot} &= h_{DL}(\hat{m}_{Py} \times (\hat{j}_e \times \hat{z})) + h_{FL}(\hat{j}_e \times \hat{z}) + h_{Oe}^{//}(\hat{j}_e \times \hat{z}) + h_{Oe}^{\perp}\hat{z} \\
&+ h_{DL}^{R}\hat{m}_{Py} \times ((\hat{j}_e \times \hat{z}) \times \hat{m}) + h_{FL}^{R}(\hat{j}_e \times \hat{z}) \times \hat{m} \\
&+ h_{DL}^{F}(\hat{m}_{Py} \times \hat{m})\left[(\hat{j}_e \times \hat{z}) \cdot \hat{m}_{PML}\right] + h_{FL}^{F}\hat{m}\left[(\hat{j}_e \times \hat{z}) \cdot \hat{m}_{PML}\right] \\
&+ \vec{H}_{ext} - M_{eff}\hat{z}(\hat{m}_{Py} \cdot \hat{z})
\end{aligned} \quad (S1)$$

where $h_{FL}$ is the effective field of field-like torque due to a spin current with conventional symmetry, $h_{Oe}^{//}$ is the in-plane Oersted field, $h_{FL}^{R}$ is the effective field of field-like torque due to spin rotation, $h_{DL}^{F}$ is the effective field of damping-like torque due to spin filtering and $h_{FL}^{F}$ is the effective field of field-like torque due to spin filtering. In equilibrium, the magnetization must satisfy the condition $\hat{m}_{Py} \times \vec{H}_{tot} = 0$.

Similarly, the PML is also subject to current-induced spin-orbit torques. A full solution to the coupled equations of motion requires numerical methods. Here we make the approximation that all terms proportional to current are treated as perturbations. Therefore, at equilibrium without an electric current, the Py magnetization is aligned along $H_{ext}$, and the PML magnetization is tilted from perpendicular toward $H_{ext}$ with an angle $\theta_{PML} = \sin^{-1}(H_{ext}/H_{an\perp})$.

Since $H_{ext}$ is along the x-axis, and current flow is along angle $\phi$ relative to the x-axis, we can solve for the equilibrium magnetization orientation in spherical coordinates as

$$\delta\theta = \frac{h_{DL}\cos\phi - h_{Oe}^\perp - h_{DL}^R\cos\theta_{PML}\sin\phi - h_{FL}^R\sin\theta_{PML}\cos\phi + h_{FL}^F\sin\theta_{PML}\cos\theta_{PML}\sin\phi}{H_{ext} + M_{eff}}$$
$$\delta\phi = \frac{-h_{Oe}^{//}\cos\phi - h_{FL}\cos\phi - h_{DL}^R\sin\theta_{PML}\cos\phi + h_{FL}^R\cos\theta_{PML}\sin\phi + h_{DL}^F\sin\theta_{PML}\cos\theta_{PML}\sin\phi}{H_{ext}}$$
(S2)

By assuming $H_{ext} \ll H_{an\perp}$ in the $\phi = 0°$ configuration, Eq. (S2) can be approximated as

$$\delta\theta \cong \frac{-h_{Oe}^\perp \pm h_{DL} \mp h_{FL}^R H_{ext}/H_{an\perp}}{H_{ext} + M_{eff}}$$
$$\delta\phi \cong \frac{-h_{Oe}^{//} - h_{FL} \mp h_{DL}^R H_{ext}/H_{an\perp}}{H_{ext}}$$
, (S3)

where the $\pm$ sign indicates the dependence of the corresponding term on the Py magnetization direction.

In the $\phi = 90°$ configuration, Eq. (S2) can be approximated as

$$\delta\theta \cong \frac{-h_{Oe}^\perp \pm h_{DL}^R \mp h_{FL}^F H_{ext}/H_{an\perp}}{H_{ext} + M_{eff}}$$
$$\delta\phi \cong \frac{\pm h_{FL}^R \pm h_{DL}^F H_{ext}/H_{an\perp}}{H_{ext}}$$
, (S4)

It should be pointed out that although the spin filtering effect in the PML can give rise to effects with a similar symmetry to that of the observed SOT, the spin filtering signal is in proportion to $H_{ext}/H_{an\perp}$, which is a higher order effect than what we observe.

## 4. Field-like SOT with spin rotation symmetry

Besides the damping-like torque, a spin current can also exert a field-like torque on a magnetic layer, particularly when the magnetic layer is thin[2]. In our measurement geometry, both $\vec{Q}_{\hat{\sigma}}$ and $\vec{Q}_{\hat{\sigma}}^R$ can therefore exert in-plane effective fields on the Py magnetization, which then rotate the Py magnetization in-plane, as described in the second equations in Eq. (S3) and (S4).

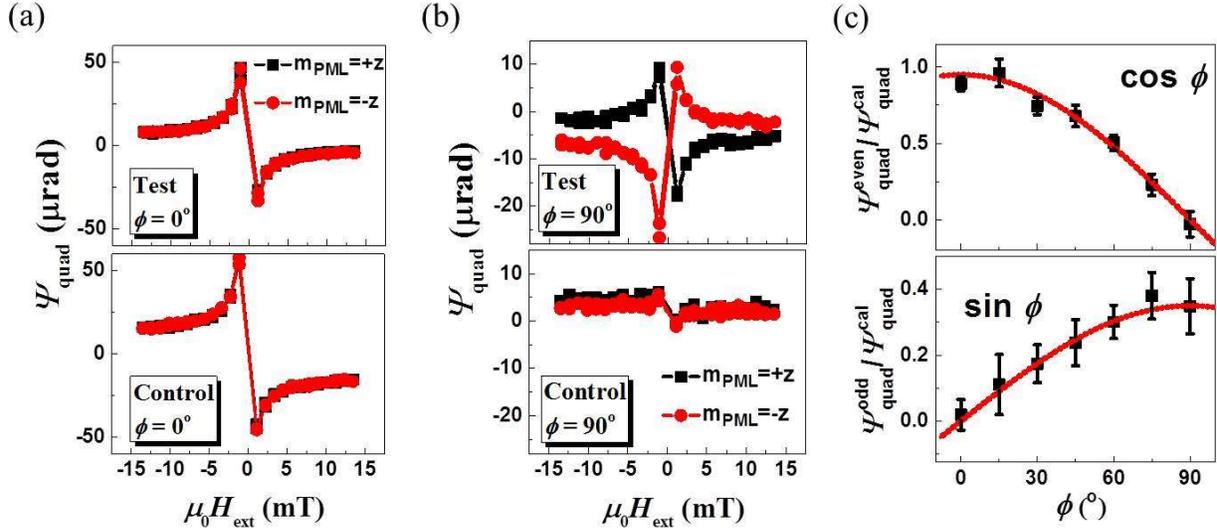

Figure S2. Current-induced in-plane effective fields measured in the same configuration as shown in Fig. 2 (a) in the main text, detected by quadratic MOKE. In this measurement, normal incidence light with linear polarization along the x-direction is used. Although in this measurement, the polar MOKE signal is also detected, it is much weaker than the quadratic MOKE response and can be distinguished from fittings. (a) The quadratic MOKE response measured in the test (top) and control samples (bottom) when current is applied parallel to $H_{ext}$. No dependence on the initial magnetization of the PML is observed. (b) The quadratic MOKE response measured in the test (top) and control samples (bottom) when current is applied perpendicularly to $H_{ext}$. In the test sample, the quadratic MOKE response is reversed when the initial magnetization of the PML is reversed. On the contrary, the weak quadratic MOKE response in the control sample, which is likely due to small misalignment of $H_{ext}$, has little dependence on the initial magnetization of the PML. (c) The angle dependence of $\psi_{quad}^{even}$ and $\psi_{quad}^{odd}$ of the test sample. Here the values are extracted by performing linear fitting with $\psi_{quad}^{cal}$, which is the quadratic MOKE response measured with an external AC calibration field of $117 \pm 10$ A/m. The slope is extracted and plotted here as $\psi_{quad}^{even}/\psi_{quad}^{cal}$ and $\psi_{quad}^{odd}/\psi_{quad}^{cal}$, respectively. Red lines are fittings using $\cos(\phi)$ and $\sin(\phi)$ functions.

We measured the in-plane reorientation of the Py magnetization by use of the quadratic MOKE (q-MOKE)[3]. We observed signals with a symmetry similar to that shown in Fig. 2 when we used the polar MOKE(p-MOKE) to detect tilting of the Py magnetization out of the film plane. The q-MOKE response $\Psi_{quad}$ is proportional to $(\hat{m}_{Py} \cdot \hat{y})(\hat{m}_{Py} \cdot \hat{x})$. When the charge current is parallel to $H_{ext}$ ($\phi = 0$, as depicted in Fig. 2(a)), shown in Fig. S2 (a), both the test and control samples exhibit q-MOKE responses that are proportional to $1/H_{ext}$ but independent of $\hat{m}$, as expected from Eq. (S3). When the charge current is applied perpendicular with $H_{ext}$ ($\phi = 90°$), as shown in Fig. S2 (b), a significant q-MOKE signal with a sign dependence on $\hat{m}$ is obtained with the test sample. This is consistent with the second equation in Eq. (S4). On the other hand, the q-MOKE signal for the control sample is independent of $\hat{m}$, as expected since $\vec{Q}_{\hat{\sigma}}^{R}$ is blocked by the TaO$_x$ layer. We measured $h_{FL}^R$ = 41 ± 13 A/m via q-MOKE with the test sample, where we employed the same current (30 mA through a 50 μm strip) as was used in the p-MOKE measurement. As such, the field-like torque is comparable in magnitude to the damping-like torque due to $\vec{Q}_{\hat{\sigma}}^R$.

We decompose $\Psi_{quad}$ into components that are even or odd in $\hat{m}$ by use of

$$\psi_{quad}^{even} = [\psi_{quad}^+ + \psi_{quad}^-]/2 \\ \psi_{quad}^{odd} = [\psi_{quad}^+ - \psi_{quad}^-]/2 \quad , \tag{S5}$$

where the superscripts + and - denotes whether $\hat{m}$ is oriented along +z or –z, respectively. We further perform linear fitting of $\psi_{quad}^{even}$ and $\psi_{quad}^{odd}$ with a calibration signal $\psi_{quad}^{cal}$ measured with an external calibration field[3], and plot the slopes as a function of $\phi$ in Fig. S2 (c). As expected from the second equation in Eq. (S2), $\psi_{quad}^{even}$, which presumably results from the sum of the in-plane Oersted field and the field-like torque generated by $\vec{Q}_{\hat{\sigma}}$, is proportional to cos($\phi$); while $\psi_{quad}^{odd}$, which is ostensibly the result of the field-like torque generated by $\vec{Q}_{\hat{\sigma}}^R$, is proportional to sin($\phi$).

## 5. Discussion of an alternative mechanism for spin rotation

In the ϕ = 90° configuration of the p-MOKE measurement, we have measured a spin current with rotated spin polarization, which we attribute to spin-orbit effects near the PML/Cu interface. However, an identically polarized spin current may be generated by a combination of the anomalous Hall effect (AHE) in Py and rotation of the resultant spin accumulation at the PML/Cu interface. In the ϕ = 90° configuration, both Pt and Py can generate a spin current with the spin polarization collinear with the Py magnetization. In the case of Pt, the spin Hall effect gives rise to the collinear spin current. In the case of Py, it is the AHE that is the ultimate source of such a collinear spin current. The resultant spin accumulation in the Cu can generate an orthogonally polarized spin current via the imaginary part of spin-mixing conductance at the PML/Cu interface. This process, as shown in Fig. S3 (a), is analogous to that discussed in the anomalous Hall-like effect in YIG/Pt bilayers[4]. However, as we discuss below, it is highly unlikely that this alternative mechanism is the dominant source of the experimentally observed SOT with spin rotation symmetry.

First, we found that the Pt capping layer plays a negligible role in the observed SOT with spin rotation symmetry. For a sample where Ta is substituted for Pt as a capping layer, i.e. seed/PML/Cu(3)/Py(2)/Ta(3), the p-MOKE signal measured in the ϕ = 0° configuration is opposite in sign to that measured from the test sample with the Pt cap, as shown in Fig. S3 (b). One possible explanation is that the spin Hall angle of Ta is opposite in sign to that of Pt, as previously reported[5]. In addition, it is also possible that the PML generates a spin current $\vec{Q}_{\hat{\sigma}}$, which generates a SOT on Py opposite to the SOT from Pt. However, shown in Fig. S3 (b), the p-MOKE signal measured at $\phi = 90°$ has the same sign as that measured from the test sample with Pt capping (Fig. 2(c) in the main text), suggesting Pt and Ta are not the main source for $\vec{Q}_{\hat{\sigma}}^{R}$ observed experimentally. In fact, the effective damping-like field due to $\vec{Q}_{\hat{\sigma}}^{R}$ measured in the sample with Ta capping is about $53 \pm 6$ A/m when applied the same total current (30 mA) as the Pt capped sample, the efficiency of which is larger than that measured in the test sample with Pt capping. We think the larger signal in Ta capped sample may be due to a slightly thinner magnetic Py layer due to possible dead layers[6], and slightly higher current density through the PML ($4.6 \times 10^{10}$ A/m$^2$) due to less shunting from the capping layer.

Secondly, we argue the spin current generated by the AHE in Py is also unlikely to be the source for the observed spin current with spin rotation. From the magnetoelectronic circuit theory for non-collinear spins[7,8], the boundary condition for the spin-accumulation and spin current at the PML/Cu interface is given by,

$$\begin{pmatrix} \vec{Q}_{\hat{x}} \\ \vec{Q}_{\hat{y}} \end{pmatrix}_{PML/Cu} = \begin{pmatrix} \text{Re}[G^{\uparrow\downarrow}] & -\text{Im}[G^{\uparrow\downarrow}] \\ \text{Im}[G^{\uparrow\downarrow}] & \text{Re}[G^{\uparrow\downarrow}] \end{pmatrix} \begin{pmatrix} \mu_{\hat{x}} \\ \mu_{\hat{y}} \end{pmatrix}_{PML/Cu}, \tag{S6}$$

where $\mu_{\hat{x}}$ and $\mu_{\hat{y}}$ are the transverse spin chemical potentials at the PML/Cu interface (decomposed in an arbitrary basis), $\vec{Q}_{\hat{x}}$ and $\vec{Q}_{\hat{y}}$ are the transverse spin currents. Here

$$G^{\uparrow\downarrow} = \frac{e^2}{2\pi\hbar} \int_{FS} \frac{d^2k}{(2\pi)^2} \left(1 - R_{\uparrow}^* R_{\downarrow}\right), \tag{S7}$$

is the interfacial spin mixing conductance at the PML/Cu interface, where the integral is over the Fermi surface (FS)[9,10], $R_{\uparrow}$ and $R_{\downarrow}$ are respectively the reflection coefficients of spin up and spin down electrons. Due to strong dephasing, it is generally believed that $\int_{FS} d^2k R_{\uparrow}^* R_{\downarrow} \ll \int_{FS} d^2k$, and therefore, $\text{Im}[G_{PML/Cu}^{\uparrow\downarrow}] \ll \text{Re}[G_{PML/Cu}^{\uparrow\downarrow}]$, which is also confirmed by first principle calculations of various FM/Cu interfaces[10]. In addition, spin-pumping theory posits that the imaginary part of the spin-mixing conductance should cause an interfacial renormalization of the gyromagnetic ratio in the case of ferromagnetic resonance[11]. Precise ferromagnetic resonance measurements of the gyromagnetic ratio as a function of Pt thickness for Py/Cu/Pt multilayers failed to find any evidence for such a renormalization[12]; strongly suggesting that $\text{Im}[G_{PML/Cu}^{\uparrow\downarrow}]$ is indeed negligible for the specific case of FM/Cu interfaces.

The upper bound for the spin current with spin rotation, which is generated by the combinational effects of the AHE in Py and the $\text{Im}[G^{\uparrow\downarrow}]$ at the PML/Cu interface, can be estimated as

$$\vec{Q}^R_{\hat{\sigma}} = \frac{\hbar}{2e} \frac{\text{Im}[G^{\uparrow\downarrow}]}{\text{Re}[G^{\uparrow\downarrow}]} \theta_{\text{Py}} j_{\text{Py}}, \qquad (S8)$$

where $\theta_{\text{Py}}$ is the effective spin Hall angle due to the AHE, $j_{\text{Py}}$ is the charge current density through Py. Here we neglect the fact that Py is thinner than its spin diffusion length[13,14], which leads to an overestimation of $\vec{Q}^R_{\hat{\sigma}}$. Using bulk conductivity $\sigma_{\text{Py}} = 4 \times 10^6 \, \Omega^{-1}\text{m}^{-1}$ [15], we estimate the upper bound of current density, $j_{\text{Py}} \approx 2 \times 10^{10} \, \text{Am}^{-2}$. Using values in literatures, $\theta_{\text{Py}} = 0.02$ [16,17] and assuming that the spin mixing conductance of the PML/Cu interface is similar to the disordered interface of Co/Cu, $\text{Re}[G^{\uparrow\downarrow}_{\text{PML/Cu}}] = 0.55 \times 10^{15} \, \Omega^{-1}\text{m}^{-2}$, $\text{Im}[G^{\uparrow\downarrow}_{\text{PML/Cu}}] = 0.03 \times 10^{15} \, \Omega^{-1}\text{m}^{-2}$ [10], we can estimate the effective field of the damping-like torque on the 2nm Py generated in this process, $h^R_{\text{DL}} = \frac{\vec{Q}^R_{\hat{\sigma}}}{\mu_0 M_{s\_\text{Py}} d_{\text{Py}}} = 3.6 \, \text{A/m}$, where $\mu_0 M_{s\_\text{Py}} = 1\text{T}$ and $d_{\text{Py}} = 2\text{nm}$. This value, though overestimated, is still an order smaller than the observed effect.

Therefore, we conclude the combinational effect of the AHE in Py and the imaginary part of spin mixing conductance at the PML/Cu interface is not the main source for the observed $\vec{Q}^R_{\hat{\sigma}}$ with spin rotation symmetry.

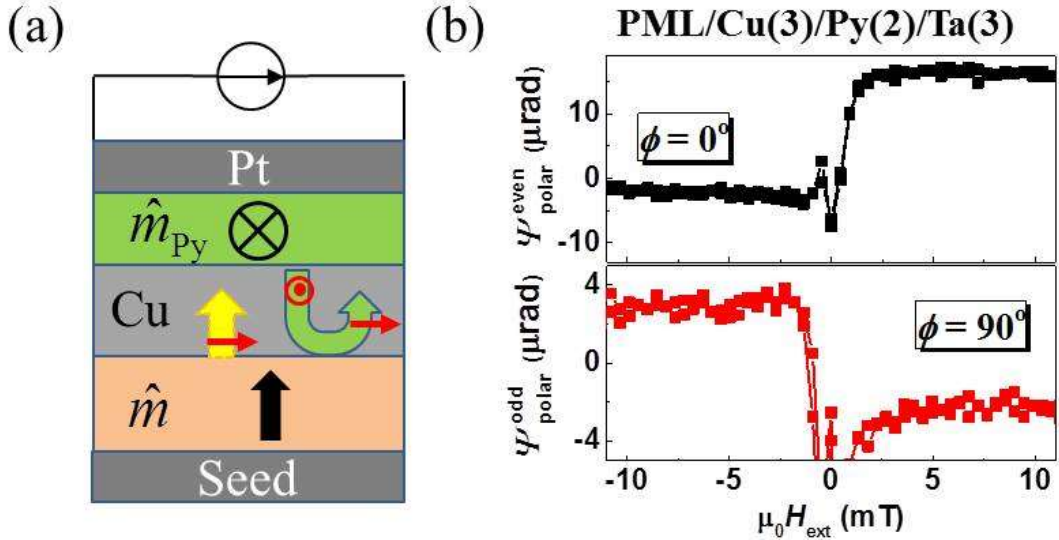

Figure S3. (a) Illustration of the alternative mechanisms that can give rise to a spin current with a rotated polarization. The mechanism that is most consistent with the measured data is depicted with the red-

yellow arrow pair, where the yellow arrow is the flow direction for spins polarized along the red arrow. In this mechanism, the spin current with rotated spins originates near the PML/Cu interface. Alternatively, the Pt and Py layers can generate a longitudinal spin accumulation in the Cu layer, which then couples to an orthogonally polarized spin current via the spin-mixing conductance at the PML/Cu interface. This mechanism is depicted by the green arrow with the red arrow denoting the spin direction. (b) p-MOKE measurement of the damping-like torque in the PML/Cu(3)/Py(2)/Ta(3) sample, i.e. the sample with the Ta cap. Top panel: p-MOKE response with an even dependence on $\hat{m}$ measured for $\phi = 0$. Bottom panel: p-MOKE response that is odd with $\hat{m}$ measured for $\phi = 90°$. Similar to Eq. (S5), $\psi_{polar}^{even}$ and $\psi_{polar}^{odd}$ are defined as $\psi_{polar}^{even} = [\psi_{polar}^{+} + \psi_{polar}^{-}]/2$, $\psi_{polar}^{odd} = [\psi_{polar}^{+} - \psi_{polar}^{-}]/2$, where the superscript denotes the sign of $\hat{m}$.

## 6. Artifacts due to the anomalous Hall effect

In both measurements of the SOT and SGE, there are potential artificial signals that have the same symmetry with $\hat{m}$, simply due to the AHE in the PML that bends the electric current. The artificial signals are expected to be very weak and can be calibrated from the control measurement with the insertion of TaO$_x$. Here we use the spin Seebeck effect-driven SGE measurement to show the origin of this artifact and the estimated order of magnitude.

As shown in Fig. S4 (a), which is equivalent to Fig. 3(a) in the main text when $\phi = 90°$, the perpendicular temperature gradient generates a voltage along the x-direction due to the SGE with conventional symmetry. In general, the SGE are different in each layers, but the equilibrium voltage at the ends of each film are the same. Therefore, even though the total electric current flowing in the film is zero, the electric current flowing in each layer is non-zero. The electric current flowing in the PML will then generate a voltage along the y-direction due to the AHE of the PML. Such a signal depends on the direction of the magnetization of both the Py and PML, which has the same symmetry as the observed SGE with spin rotation symmetry. We model such an artificial signal with a parallel circuit model, as shown in Fig. S4 (b) and (c). The multilayers are separated into three regimes: the first regime includes the Py and Pt layer, which contributes to the SGE with conventional symmetry; the second regime includes the PML, which contributes to the AHE; and the third regime that includes all other layers, which shunts the charge current.

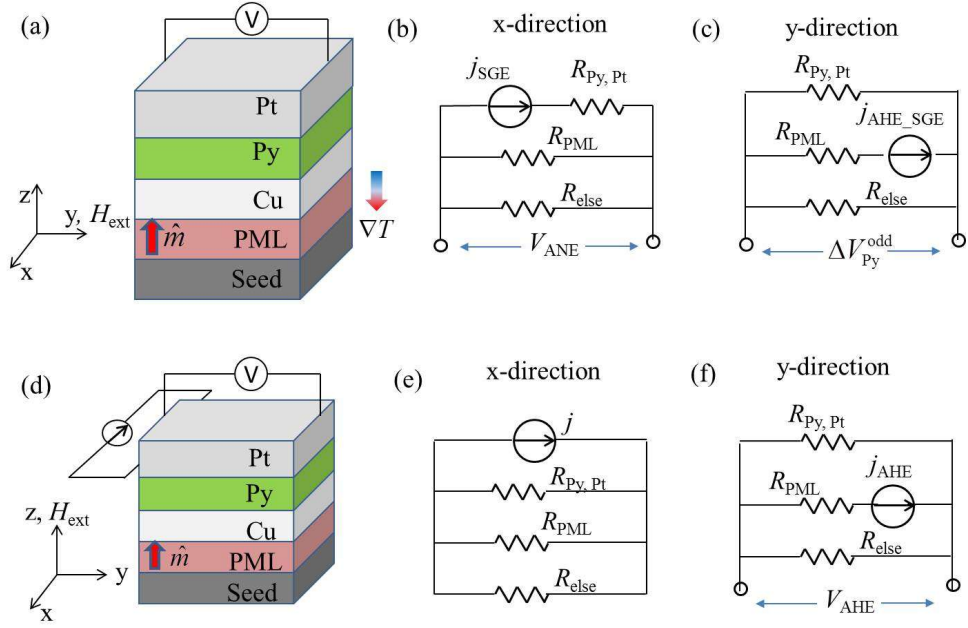

Figure S4. (a) Configuration where the SGE with spin rotation symmetry is observed. (b) Circuit diagram in the x-direction driven by the SGE with conventional symmetry. (c) Circuit diagram in the y-direction driven by the AHE in the PML. (d) Configuration of an AHE measurement. (e) Circuit diagram in the x-direction driven by the applied current. (f) Circuit diagram in the y-direction driven by the AHE in the PML.

The effective current density in each regime shall satisfy

$$\begin{bmatrix} j_{1x} \\ j_{1y} \end{bmatrix} = \begin{bmatrix} \sigma_1 & 0 \\ 0 & \sigma_1 \end{bmatrix} \begin{bmatrix} E_x \\ E_y \end{bmatrix} + \begin{bmatrix} j_{SGE} \\ 0 \end{bmatrix}$$

$$\begin{bmatrix} j_{2x} \\ j_{2y} \end{bmatrix} = \begin{bmatrix} \sigma_2 & -\sigma_2^{AH} \\ \sigma_2^{AH} & \sigma_2 \end{bmatrix} \begin{bmatrix} E_x \\ E_y \end{bmatrix}$$

$$\begin{bmatrix} j_{3x} \\ j_{3y} \end{bmatrix} = \begin{bmatrix} \sigma_3 & 0 \\ 0 & \sigma_3 \end{bmatrix} \begin{bmatrix} E_x \\ E_y \end{bmatrix}$$

$$\begin{bmatrix} j_{1x} \\ j_{1y} \end{bmatrix} d_1 + \begin{bmatrix} j_{2x} \\ j_{2y} \end{bmatrix} d_2 + \begin{bmatrix} j_{3x} \\ j_{3y} \end{bmatrix} d_3 = 0$$

(S9)

where $\sigma$ and $d$ are respectively the effective conductivity and thickness of the three regimes with the subscript denoting the corresponding regime, $\sigma_2^{AH}$ is the anomalous Hall conductivity of the PML, $j_{SGE}$ is the electric current density due to the SGE in Py and Pt with conventional symmetry. The last equation is due to the open circuit boundary condition, where the total currents in both

the x-direction and y-direction vanish. The relation between $E_x$ and $E_y$ can be derived from Eq. (S9) as

$$\frac{E_y}{E_x} = \frac{\sigma_2^{AH} d_2}{\sigma_1 d_1 + \sigma_2 d_2 + \sigma_3 d_3}. \tag{S10}$$

The value shown in Eq. (S10) happens to be the total anomalous Hall angle in the same sample that can be determined by the Hall measurement. As shown in Fig. S4 (d), in a typical Hall measurement, the current distribution in the x- and y-directions can be calculated as

$$\begin{bmatrix} j_{1x} \\ j_{1y} \end{bmatrix} = \begin{bmatrix} \sigma_1 & 0 \\ 0 & \sigma_1 \end{bmatrix} \begin{bmatrix} E_x \\ E_y \end{bmatrix}$$
$$\begin{bmatrix} j_{2x} \\ j_{2y} \end{bmatrix} = \begin{bmatrix} \sigma_2 & -\sigma_2^{AH} \\ \sigma_2^{AH} & \sigma_2 \end{bmatrix} \begin{bmatrix} E_x \\ E_y \end{bmatrix}$$
$$\begin{bmatrix} j_{3x} \\ j_{3y} \end{bmatrix} = \begin{bmatrix} \sigma_3 & 0 \\ 0 & \sigma_3 \end{bmatrix} \begin{bmatrix} E_x \\ E_y \end{bmatrix}$$
$$\begin{bmatrix} j_{1x} \\ j_{1y} \end{bmatrix} d_1 + \begin{bmatrix} j_{2x} \\ j_{2y} \end{bmatrix} d_2 + \begin{bmatrix} j_{3x} \\ j_{3y} \end{bmatrix} d_3 = \begin{bmatrix} K_{tot} \\ 0 \end{bmatrix} \tag{S11}$$

where $K_{tot}$ is the total sheet current applied. Therefore, the total anomalous Hall angle can be calculated as

$$\theta_{AHE\_tot} = \frac{E_y}{E_x} = \frac{\sigma_2^{AH} d_2}{\sigma_1 d_1 + \sigma_2 d_2 + \sigma_3 d_3}. \tag{S12}$$

The total anomalous Hall angle is experimentally determined to be 0.02%. Therefore, the SGE signal with spin rotation, $\Delta V_{Py}^{odd}$, in the test sample (Seed/PML/Cu/Py/Pt) due to this artifact is estimated to be 4 nV–much smaller than the 0.6 μV signal observed in Fig. 3(b) in the main text. The similar artifact in the MOKE measurement also scales with the total anomalous Hall angle, and is thus negligible compared to the observed signals.

### 7. Artifacts due to the interlayer magneto static coupling

If there is an interlayer magneto static coupling between the PML and Py layers, the magnetizations of the PML and Py will be slightly tilted away from the designated directions. This can also potentially cause artificial signals in the measurements of SOT and SGE with spin rotation symmetry. In our sample, the interlayer coupling is likely to be dominated by orange peel coupling, of which the effective field is usually about 2 mT or less[18,19]. We estimate the artificial signals due to interlayer coupling to be at least an order smaller than our measured signal.

In the measurement of SOT with spin rotation symmetry, the orange peel coupling may give rise to magnetization tilting as shown in Fig. S5. We assume the interlayer coupling field applied on the in-plane magnetized Py is $H_{\text{Py}}^{\text{ILC}}\hat{m}$, and that applied on the PML is $H_{\text{PML}}^{\text{ILC}}\hat{m}_{\text{Py}}$. Here we assume $H_{\text{Py}}^{\text{ILC}} \approx H_{\text{PML}}^{\text{ILC}} \approx 2\text{mT}$. Taking the approximation that the interlayer coupling field is small, the magnetization tilting in each layer can be calculated as

$$\left(H_{\text{Py}}^{\text{ILC}}\hat{m}\cdot\hat{z} - M_{\text{eff}}\sin\theta_{\text{Py}}\right)\cos\theta_{\text{Py}} = \left(h_{\text{Py}}^{//} + H_{\text{ext}} + H_{\text{an}}\cos\theta_{\text{Py}}\right)\sin\theta_{\text{Py}}$$
$$H_{\text{an}\perp}\cos\theta_{\text{PML}}\sin\theta_{\text{PML}} = \left(h_{\text{PML}}^{//} + H_{\text{ext}} + H_{\text{PML}}^{\text{ILC}}\hat{m}_{\text{Py}}\cdot\hat{x}\right)\cos\theta_{\text{PML}}$$
, (S13)

where $H_{\text{an}}$ is the in-plane anisotropy of Py, $\mu_0 M_{\text{eff}} = 0.74$ T and $\mu_0 H_{\text{an}\perp} = 0.39$ T are obtained from ferromagnetic resonance measurements, $h_{\text{Py}}^{//}$ and $h_{\text{PML}}^{//}$ are the effective fields applied to Py and the PML, respectively, due to the combinational effect of current-induced Oersted field and field-like torque. Under the reasonable approximation that $h_{\text{Py}}^{//}$, $h_{\text{PML}}^{//} \ll H_{\text{ext}}$, $H_{\text{Py}}^{\text{ILC}}$, $H_{\text{PML}}^{\text{ILC}} \ll M_{\text{eff}}$, $H_{\text{an}\perp}$, we can solve Eq. (S13),

$$\theta_{\text{Py}} \approx \frac{H_{\text{Py}}^{\text{ILC}}(\hat{m}\cdot\hat{z})}{\left(h_{\text{Py}}^{//} + H_{\text{ext}} + H_{\text{an}}\right)(\hat{m}_{\text{Py}}\cdot\hat{x}) + M_{\text{eff}}}$$
$$\theta_{\text{PML}} \approx \frac{h_{\text{PML}}^{//} + H_{\text{ext}} + H_{\text{PML}}^{\text{ILC}}(\hat{m}_{\text{Py}}\cdot\hat{x})}{H_{\text{an}\perp}}$$
. (S14)

Therefore, the perturbation from $h_{\text{Py}}^{//}$ and $h_{\text{PML}}^{//}$ to the out-of-plane magnetization reorientation is

$$\Delta m_{\text{Py}}^{\perp} \approx \Delta\theta_{\text{Py}} \approx \frac{H_{\text{Py}}^{\text{ILC}}(\hat{m}\cdot\hat{z})}{M_{\text{eff}}^2}h_{\text{Py}}^{//}(\hat{m}_{\text{Py}}\cdot\hat{x})$$
$$\Delta m_{\text{PML}}^{\perp} \approx (\hat{m}\cdot\hat{z})\sin\theta_{\text{PML}}\Delta\theta_{\text{PML}} \approx \frac{H_{\text{ext}} + H_{\text{PML}}^{\text{ILC}}(\hat{m}_{\text{Py}}\cdot\hat{x})}{H_{\text{an}\perp}^2}h_{\text{PML}}^{//}(\hat{m}\cdot\hat{z})$$
. (S15)

Through a symmetry-based analysis, one can find that both $\Delta m_{\text{Py}}^{\perp}$ and $\Delta m_{\text{PML}}^{\perp}$ in Eq. (S15) change as the magnetization of either the Py or PML switches, which has the same symmetry as the observed out-of-plane magnetization reorientation due to the rotated spin-orbit torques. The artificial signal is only due to the in-plane current-induced fields. However, we expect these artifacts to have either significantly lower signals than the observed values or to have a different line shape from the observed external field dependence. For example, the first equation of Eq. (S15) describes how Py will have a similar line shape and symmetry as the effect from the rotated spin-orbit torques. However, when compared with the formula of the latter, which is $\Delta m_{\text{Py}}^{\perp} \approx \frac{1}{M_{\text{eff}}}h_{\text{DL}}^{\text{R}}$, the former has an additional scaling factor of $\frac{H_{\text{Py}}^{\text{ILC}}}{M_{\text{eff}}}$, which is estimated to be

0.3%. From Fig. S2(a), we estimate $h_{Py}^{//}$ to be about 100 A/m. Multiplying this by 0.3% gives rise to 0.3 A/m, which is two orders smaller than what is required to achieve the $\Delta m_{Py}^{\perp}$ observed experimentally ($h_{DL}^{R}$ = 30 ± 4 A/m). Moreover, we have performed a control measurement by applying an external oscillating magnetic field of 206 A/m to emulate the possible effect due to $h_{Py}^{//}$. However, to our measurement accuracy, we did not observe any signal as those shown in Fig. 2. Therefore, we conclude the artificial signal due to the slight out-of-plane Py magnetization tilting, which may appear to have the same symmetry as the rotated SOT, is negligible.

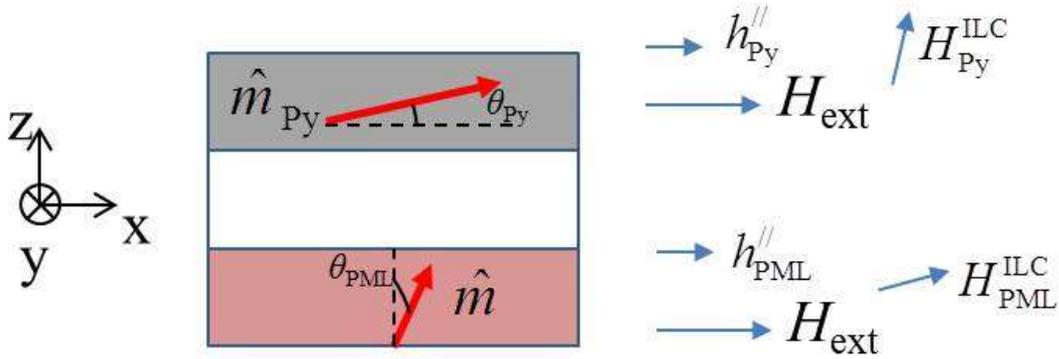

Figure S5. Illustration of magnetization reorientation due to interlayer magneto static coupling. In the SOT with spin rotation symmetry measurement, the charge current is applied along the y-direction. In the SGE with spin rotation symmetry measurement, the voltage is measured in the x-direction. The fields acting on the Py (top three vectors) and PML (bottom three vectors) layers are sketched on the right hand side. For simplicity, the demagnetizing field and anisotropy field are not plotted.

Similarly, the second equation of Eq. (S15) also has an additional scaling factor $\frac{H_{PML}^{ILC}}{H_{an\perp}} \approx 0.5\%$, therefore we expect that the artificial signal due to this effect is also weak. In addition, the second equation of Eq. (S15) suggests that the polar MOKE response should also carry a linear dependence with $H_{ext}$, which is hardly observed in the experiments. If we use the second equation of Eq. (S15) to fit the observed data in Fig. 2(c) in the main text, this will lead to $H_{Py}^{ILC}$ greater than 0.1 T, which is 50 times higher than a typical orange peel coupling induced interlayer coupling field.

In addition, the magnetization tilting due to the interlayer magneto static coupling cannot explain the field-like torque with spin rotation symmetry as shown in Fig. S2. Therefore, we conclude Eq. (S15) cannot quantitatively describe the observed signals.

In the experiments of SGE with spin rotation symmetry, magnetization tilting due to interlayer coupling gives rise to additional PNE signals in the Py and PML layers that are proportional to $\theta_{Py}$ and $\theta_{PML}$, respectively.

$$V_{Py}^{PNE} \propto (\hat{m}_{Py} \cdot \hat{x})\theta_{Py} \approx \frac{H_{Py}^{ILC}(\hat{m} \cdot \hat{z})(\hat{m}_{Py} \cdot \hat{x})}{M_{eff}}$$

$$V_{PML}^{PNE} \propto (\hat{m} \cdot \hat{z})\theta_{PML} \approx \frac{H_{ext}(\hat{m} \cdot \hat{z}) + H_{PML}^{ILC}(\hat{m}_{Py} \cdot \hat{x})(\hat{m} \cdot \hat{z})}{H_{an\perp}}. \quad (S16)$$

The first term of the second equation in Eq. (S16) gives rise to the PNE in the PML, which is in fact, observed as the straight line in the bottom panel of Fig. 3(b). The second term of the second equation in Eq. (S16) resembles the symmetry of $\Delta V_{Py}^{odd}$ as observed in Fig. 3 in the main text. However, fitting the black curve in the bottom panel of Fig. 3 (b) with Eq. (S16) will give rise to a $H_{PML}^{ILC}$ as large as 40 mT, one order higher than the possible dipolar coupling field (2mT). Similarly, the first equation in Eq. (S16) also resembles the symmetry of $\Delta V_{Py}^{odd}$. However, given that Py is thinner than the PML, we expect this signal to be even smaller than the possible artifact due to the second equation in Eq. (S16).

In all, we have shown that the interlayer magneto static coupling may give rise to artificial signals, which appears to have spin rotation symmetry. However, in the material system studied here, these artificial signals are expected to be orders smaller than our measured signals.